\documentclass[12pt,preprint]{aastex}

\shorttitle{{\it{Hubble Space Telescope}} Spectroscopy of VW Hyi in Superoutburst}
\shortauthors{Merritt et al.}

\begin{document}

\title{Hubble Space Telescope Far Ultraviolet Spectroscopy of the Dwarf Nova VW Hyi in
Superoutburst 
\footnote{Based on observations made with the NASA-CNES-CSA {\it{Hubble
Space Telescope}}. {\it{Hubble}} is operated for NASA
by AURA
under NASA contract NAS5-32985}}  

\author{Jason Merritt}
\affil{Astronomy and Astrophysics, Villanova University, \\ 
800 Lancaster Avenue, Villanova, PA 19085, USA}
\email{jason.merritt@villanova.edu}

\author{Christopher Night}
\affil{Center for Astrophysics, Harvard University, 
60 Garden Street, Cambridge, MA 10038}
\email{cnight@cfa.harvard.edu}

\author{Edward M. Sion}
\affil{Astronomy and Astrophysics, Villanova University, \\ 
800 Lancaster Avenue, Villanova, PA 19085, USA}
\email{edward.sion@villanova.edu}

\clearpage 

\begin{abstract}

We obtained three consecutive HST spectroscopic observations of a single
superoutburst of the dwarf nova VW Hydri. The spectra cover the beginning,
middle, and end of the superoutburst. All of the spectra are dominated by
strong absorption lines due to CIII (1175 \AA), Lyman alpha (1216 \AA), NIV
(1238 \AA, 1242 \AA), SII (1260-65 \AA), SIII (1300 \AA), CII (1335 \AA), SIV (1394
\AA, 1402 \AA) and CIV (1548 \AA, 1550 \AA). We discuss the evolution of the far
UV energy distribution and line structure during the superoutburst. We
note the absence of any P Cygni line structure in the STIS spectra.
Using state of the art accretion disk models by Wade and Hubeny, we have
determined accretion rates for all three spectra, for two white dwarf
masses, 0.55 M$_{\sun}$ and 0.8 M$-{\sun}$. For both white dwarf masses the accretion
rate during superoutburst decreased by a factor of two from early to late
in the superoutburst. The average accretion rate during superoutburst is $3
- 6 \times 10^{-9}$M$_{\sun}$/yr depending on the white dwarf mass.

\end{abstract}


\section{Introduction} 

Dwarf novae (DNe) are a subclass of cataclysmic variable (CV) systems, in
which a white dwarf (WD, the primary) accretes hydrogen-rich matter from a
low-mass main sequence-like star (the secondary) filling its Roche lobe.
In these systems, the transferred gas forms an accretion disk around the
WD.  It is believed that the accretion disk is subject to a
thermal-viscous instability that causes cyclic changes of the accretion
rate. A low rate of accretion ($\approx 10^{-11} M_{\odot}$~yr$^{-1}$)
quiescent stage is followed every few weeks to months by a high rate of
accretion ($\approx 10^{-8} M_{\odot}$~yr$^{-1}$) outburst stage of days
to weeks. These outbursts (dwarf nova - DN accretion event or nova-like
high state), are believed to be punctuated every few thousand years or
more by a thermonuclear runaway (TNR) explosion: the classical nova
\citep{hac93}.

VW Hyi is a key system for understanding DNe in general.  It is one of the
closest \citep[placed it at 65 pc]{war87}, brightest and best-studied example of an SU
UMa-type DN and it lies along a line of sight with an exceptionally low
interstellar column \citep[estimated the HI column to be $\approx 6 \times
10^{17}$ cm$^{-2}$]{pol90}, which has permitted study of VW Hyi in nearly
all wavelength ranges, including detection in the usually opaque extreme
ultraviolet [EUVE \citep{mau96}]. Coherent and quasi-coherent soft X-ray oscillations
and a surprisingly low luminosity boundary layer (BL) have been detected
\citep{bel91,mau91}.  VW Hyi is below the CV period gap near its lower
edge, with an orbital period of 107 minutes and a quiescent optical
magnitude of 13.8. Below the period gap, gravitational wave emission is
thought to drive mass transfer, resulting in very low accretion rates
during dwarf nova quiescence.  

Systems of the SU UMa class undergo both normal DN outbursts and
superoutbursts. For VW Hyi, the normal outbursts last 1-3 days and occur
every 20-30 days, with peak visual magnitude of 9.5. The superoutbursts
last 10-15 days and occur every 5-6 months, with peak visual magnitude
reaching 8.5. The mass of the accreting WD was estimated to be 0.63
$M_{\odot}$ \citep{sch81}, but more recently a gravitational redshift
determination yielded a larger mass $M_{wd}=0.86 M_{\odot}$ \citep{sio97}.  
The inclination of the system is $\approx 60$ degrees
\citep{hua96a,hua96b}. 

While the physical properties of the accreting white dwarf in VW Hyi and
its response to heating by the outbursts and superoutbursts have been
derived from HST spectra obtained during quiescence, relatively little has
appeared on the accretion rate of VW Hyi during outbursts and
superoutbursts. In particular, the FUV spectra in outburst or
superoutburst have not as yet been compared with realistic accretion disk models with vertical structure. Note however that previous models used by \citet{hua96a} were stellar atmosphere proxies to a proper disk which nonetheless provided an excellent  approximation to vertical structure models since the disk is optically  thick, and it has been shown in the literature that for optically thick disks there is little difference between stellar atmospheres and disk-proper models.  

We have obtained a series of HST spectra of VW Hyi during a
superoutburst. In Section 2, we describe the observations and provide a
description of the spectra that were obtained.  In Section 3, we compare
the spectrum to models in an attempt to account for the continuum energy
distribution and the line spectrum. In Section 4, we discuss the results
of our synthetic spectral analysis and in Section 5, we briefly summarize
our conclusions.

\section{Observations} 

The observations of VW Hydri were obtained between 20 May 2000 and 25 May
2000, during the decline from superoutburst. For all three observations,
the instrumental setup used STIS with the FUV-MAMA detectors in ACCUM
mode. The spectra had a center wavelength of 1425A and a bandwidth of
595A, for a total wavelength range of 1140.0A to 1735A. Specific details
of the exposures are given in table 1. There were no apparent problems
with any of the spectra.

The most obvious absorption lines are NII (1085.7 \AA ), CIII
(1174.9-1176.4 \AA ), Lyman Alpha (1216 \AA), N V (1238, 1242 \AA), Si II
1260-1265 \AA), Si III + O I (1300 \AA), C II (1335 \AA), O V (1371 \AA),
Si IV (1393, 1402 \AA) and C IV (1548, 1550 \AA). None of the absorption
features exhibit P Cygni structure or obvious asymmetries which are typical signatures
signatures of wind outflow. The absorption lines gradually weaken from the
first to the third spectrum as the continuum in the relatively line free
region at 1460 \AA declined from $1.3\times 10^{-11}$ ergs/cm$^2$/s/\AA (early) to $6\times
10^{-12}$ ergs/cm$^{2}$/s/\AA (late). This is accompanied by a decrease in the
slope of the continuum from early to late in the superoutburst.

\section{Analysis}

In this section we describe the procedure we followed to assess the
contribution of the accretion disk, and derive the accretion rate onto the
white dwarf during the superoutburst. This procedure consists of comparing
the observed HST STIS spectra of VW Hyi with a grid of theoretical
accretion disk spectra for adopted system parameters and varying accretion
disk parameters. The best fit models are then obtained using a $\chi^{2}$
minimization fitting procedure. Details of the codes and the
$\chi^{2}_{\nu}$ ($\chi^2$ per degree of freedom) minimization fitting
procedures are discussed in detail in \citet{sio95} and \citet{hua96a},
and will not be repeated here.

For the accretion disk spectrum, we used the grid of accretion disk
spectra computed by \citet{wad98}, who use a slightly different version of
the code TLUSTY \citet{hub88} named TLUSDISK to generate the theoretical
spectrum of an accretion disk. The optically thick accretion disk model is
made of a collection of rings. The disk models are computed assuming LTE
and vertical hydrostatic equilibrium. The radial temperature structure in
the disk is governed by the assumption of steady state accretion
equilibrium where each annulus receives the same amount of mass as it
loses (i.e. the disk gains as much mass from the Roche lobe filling donor
star as the white dwarf accretes from the disk). Irradiation from external
sources is neglected. Local spectra of disk annuli are computed taking
into account line transitions from elements 1-28 (H through Ni). Limb
darkening as well as Doppler broadening and blending of lines are taken
into account.

Before we carried out the model fits, we masked wavelength
regions where strong resonance line absorption due to zero volt resonance
doublets appear in the HST spectra. The regions we masked in the fitting
are N V [1228-1250\AA] and C IV [1530-1552\AA] as well as any negative
fluxes. For the accretion disk fits, we "fine-tuned" the 
derived accretion rate of the best-fitting disk model by changing the 
accretion rate in increments of 0.1 over the range 0.1 to 10, on the 
assumption that the disk fluxes scale linearly over that range.

\subsection{The Accretion Disk Models}

In superoutburst, a reasonable expectation is that the far UV radiation 
should be dominated by the light of the luminous accretion disk. Moreover, 
since it is expected that at least during part of the superoutburst, the 
accretion disk should closely approximate a steady state, we explored 
whether an accretion disk model would produce better agreement with the 
{\it{HST STIS}} spectra.  Although one expects the inner region of the 
accretion disk in VW Hyi to be optically thin during quiescence, This is 
not the case during outburst or superoutburst and led us assess how well 
optically thick steady state disks can represent the observations.

In the present work we used the grid of accretion disk spectra of 
\citet{wad98} consisting of 26 different combinations of $M_{wd}$ (0.35, 
0.55, 0.80, 1.03 and 1.21 $M_{\odot}$) and $\dot{M}$ ($log \dot{M}$ = 
-8.0, -8.5, -9.0, -9.5, -10.0 and -10.5 $\dot{M}$ yr$^{-1}$; see Table 2 
in \citet{wad98}). The spectra are presented for six different disk 
inclinations $i$ (8.1, 18.2, 41.4, 60.0, 75.5 and 81.4 degrees). The 
models fluxes include the effects of limb darkening, the projection of 
fluxes as a function of the inclination angle, and are scaled to a 
distance of 100 pc where the distance is related to the scale factor as $$ 
d=100(pc)/\sqrt{S}.$$

For a system as well-studied as VW Hyi, we adopted widely used parameters
in the literature. We fixed the distance d $=65$ pc, fixed the orbital
inclination at $i = 60$ degrees but assumed two values of the white dwarf
mass, $M_{wd}$. The synthetic disk spectra were then fitted to the three
superoutburst spectra by our $\chi^{2}$ minimization routine. The
best-fitting accretion disk models are listed in Table 2 corresponding to
$M_{wd}= 0.55 M_{\odot}$ and $M_{wd}= 0.8 M_{\odot}$. We list in column
(1) the part of the superoutburst in which the spectrum was taken; (2) the
white dwarf mass adopted; (3) the adopted orbital inclination; (4) log of
the accretion rate in solar masses per year resulting from the
best-fitting disk model; (5) the $\chi^{2}$ value corresponding to the
best fit; (6) the value of the increment which produced the best refined
disk model fit (see section 3).

The best-fitting accretion disk models to the three HST spectra for both
values of the white dwarf mass are displayed in Figure 1 for the early
spectrum, Figure 2 for the middle spectrum and Figure 3 for the late
spectrum.  All three spectra are well-fit by steady state accretion disk
models except for the observed continuum shortward of Lyman Alpha where
the theoretical disk flux overpredicts the observed flux. We discuss the
results of these fits in the following section.

\section{Summary and Conclusions}

The high quality HST STIS together with the currently accepted orbital 
inclination and reasonable distance (65 pc) for VW Hydri enabled a 
determination of the accretion rate during a superoutburst as a function 
of time within the superoutburst. For assumed white dwarf masses   
$M_{wd}=0.6 M_{\odot}$ and $M_{wd}=0.8 M_{\odot}$, the derived average accretion rates during superoutburst are, 
respectively $6(\pm -1) \times 10^{-9}$ M$_{\sun}$/yr and $3(\pm -1) 
\times 10^{-9}$ M$_{\sun}$/yr. All of the best-fitting models show a flux 
excess relative to the observations. This flux deficit relative to the 
models, shortward of Lyman $\alpha$, may be due to the possibility that 
the accretion disk in VW Hydri is modestly truncated hence having a cooler inner disk region than the untruncated accretion disk model. The accretion rate late in the superoutburst as declined by a factor of two relative to the accretion rate early in the superoutburst.

\citet{hua96a} modeled an HST FOS G130H spectrum of VW Hyi taken roughly five days after the optical rise to superoutburst in October, 1993. This was approximately the same time that elapsed before 
the first STIS E140M spectrum was taken during the May, 2000 superoutburst reported in this paper. For a white dwarf mass of 
$M_{wd}=0.6 M_{\odot}$ and a distance of 90 pc, \citet{hua96a}
found a best-fitting disk model of $3(\pm -1) \times 10^{-9}$ M$_{\sun}$/yr while in the present
paper, with d = 65 pc, accretion rates of $8(\pm -1) \times 10^{-9}$ M$_{\sun}$/yr and $4(\pm -1) \times 10^{-9}$ M$_{\sun}$/yr
were determined for white dwarf masses $M_{wd}=0.6 M_{\odot}$ and $M_{wd}=0.8 M_{\odot}$, respectively. The same sharp cores of N V and C IV detected by \citet{hua96a} are seen at N V in Figures 1,2, and 3 but not C IV in this paper. In Huang et al., the sharp cores at C IV reveal a hint of inverse P Cygni structure as though either infall onto the white dwarf or superposition of an inhomogeneous accretion stream is being seen. The C IV profile in the STIS E140M spectra reported in this paper do not reveal the resolved sharp doublet cores in C IV but rather a merged sharp core. 

Given our derived accretion rates, it may be possible to carry out
quasi-static evolutionary model simulations at these rates, to compare
with the amount of white dwarf cooling evident from model analyses of
post-outburst and post-superoutburst far ultraviolet spectra. We leave 
this for a future exploration.

\acknowledgements 
This work was supported by NSF grants AST05-07514, NASA
grant NNG04GE78G and by summer undergraduate research
support from the Delaware Space Grant Consortium.

\clearpage

\begin{deluxetable}{lccccc}  
 \tablecaption{HST STIS Observations of a Superoutburst of VW Hydri}
\tablehead{
\colhead{Dataset}&\colhead{Exposure Start Time}&\colhead{Exposure Time}&\colhead{Aperture}&\colhead{Disperser}
&\colhead{  SOB Phase}
}
\startdata
O5E202010 &       2000-05-20 14:06:00  &  2512.8  &     $0.2\times 0.2$&   E140M&      Early\\
O5E203010  &      2000-05-22 17:30:00   & 2512.8   &  $  0.2\times 0.2$&   E140M &     Middle\\
O5E204010   &     2000-05-25 14:34:00    &2512.8    &   $0.2 \times 0.2$&   E140M&      Late \\
\enddata
\end{deluxetable}

\clearpage

\begin{deluxetable}{lccccc}
\tablecaption{            
Accretion Disk Model Fitting Results}
\tablehead{
\colhead{SOB Phase}
&\colhead{WD(M$_{\sun}$)}
&\colhead{$ i $}
&\colhead{log \.{M}}
&\colhead{$\chi^{2}$}
&\colhead{$ f \times \dot{M}$}
}
\startdata	   
Early  &       0.55  &      60  &     $-8.097 (8\times 10^{-9})$&          2.905 &        0.8\\
 &             0.8    &     60   &   $ -8.386 (4\times 10^{-9})$&          2.288 &       1.3\\
Middle &       0.55    &    60    &  $ -8.222 (6 \times 10^{-9})$&          2.581&         0.6\\
&              0.8      &   60   &    $-8.500 (3 \times 10^{-9})$&          2.605 &        1.0\\
Late  &        0.55      &  60    & $  -8.459 (3.5 \times 10^{-9})$&        2.950  &       1.1\\
&              0.8        & 60     &  $-8.745 (1.8\times 10^{-9})$&        3.036&         1.8            \\
\enddata
\end{deluxetable}

\clearpage

\begin{figure}
\plotone{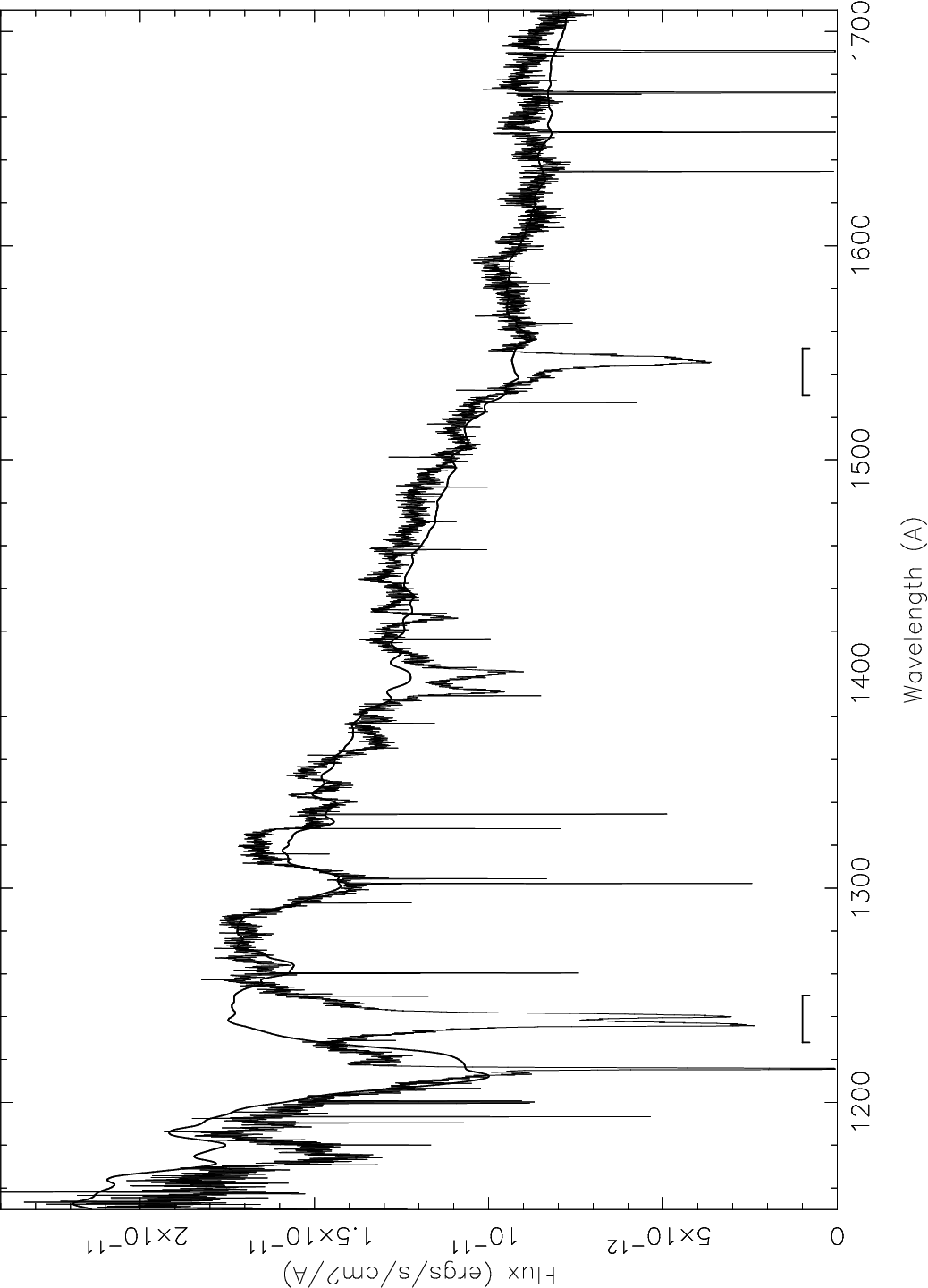}
\caption{Fig.1 - The earliest of the three spectra was taken on 20 May 2000, near 
the peak of the superoutburst. The black line shows the best fit model, 
with a mass of 0.8 M$_{\sun}$, an accretion rate of $4 \times 10^{-9}$ 
M$_{\sun}$ yr$^{-1}$, and an inclination of 60$^{o}$. The model flux shows 
an excess over the observed flux at the shortest wavelengths. The 
horizontal lines in the lower part of the figure indicate the wavelength 
regions which were masked out in the fitting.}
\end{figure}

\clearpage

\begin{figure}
\plotone{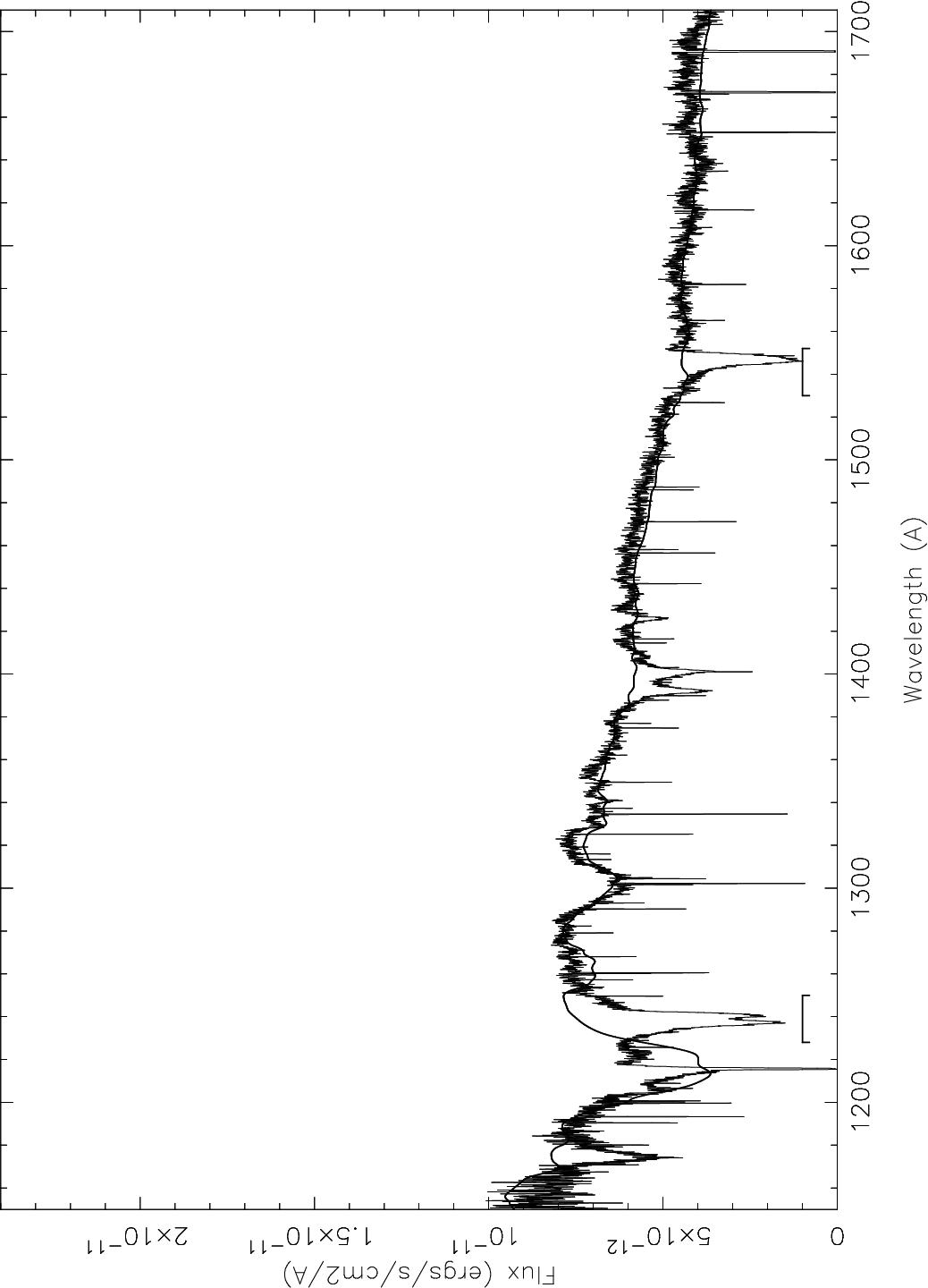}
\caption{Fig.2 - The second spectrum was taken on 22 May 2000, as the flux was 
declining after the superoutburst. The black line shows the best fit 
model, with a mass of 0.8 M$_{\sun}$, an accretion rate of $3 \times 
10^{-9}$ M$_{\sun}$ yr$^{-1}$, and an inclination of $60^{o}$.}
\end{figure}

\clearpage

\begin{figure}
\plotone{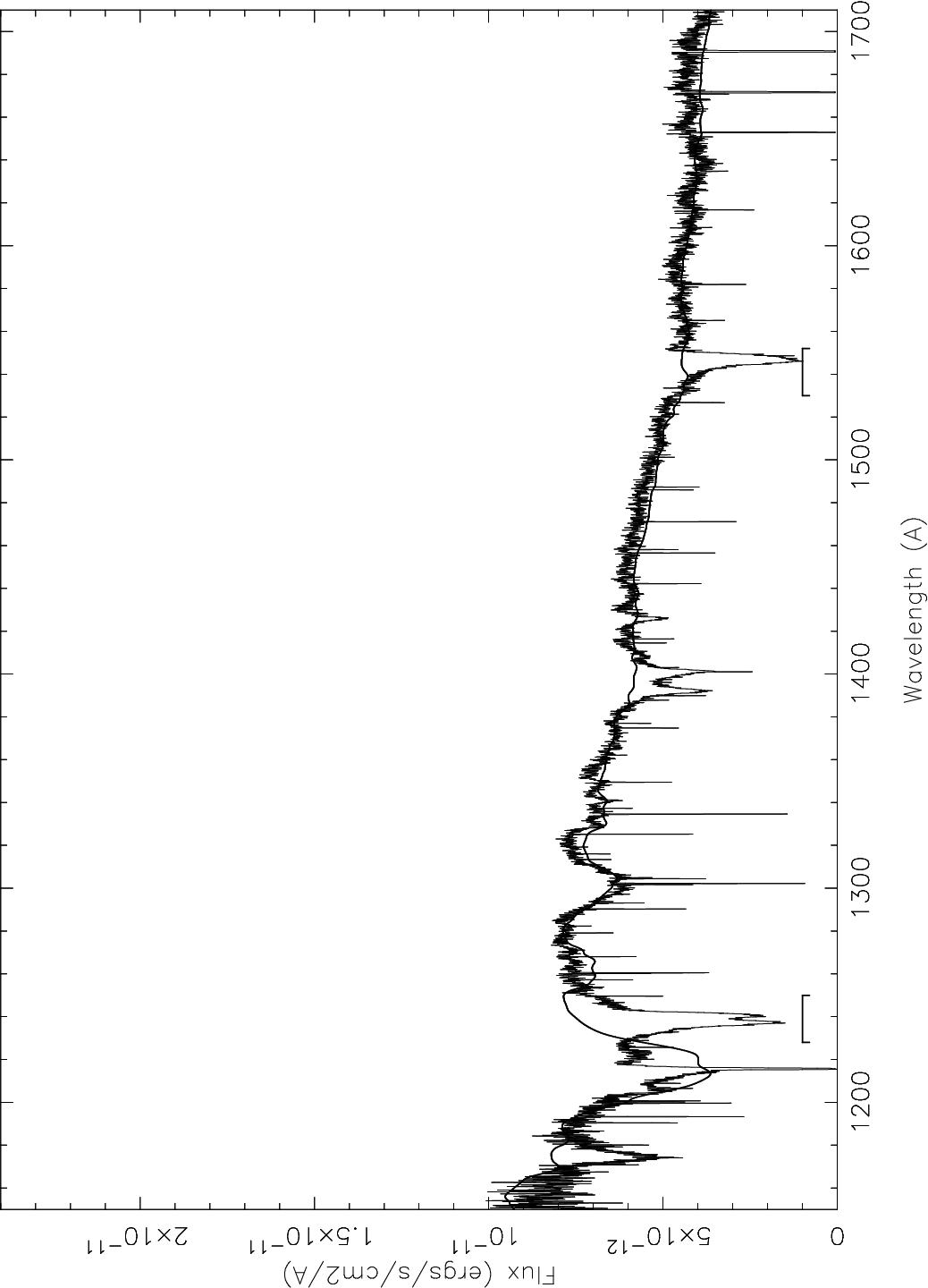}
\caption{Fig.3 - The third and final spectrum was taken on 25 May 2000, near the 
end of the superoutburst. The black line shows the best fit model, with a 
mass of 0.8 M$_{\sun}$, an accretion rate of $2\times 10^{-9}$ M$_{\sun}$ 
yr$^{-1}$, and an inclination of $60^{o}$. Toward the end of the 
superoutburst, the accretion rate has declined by a factor of 2.}
\end{figure}

\end{document}